\documentclass[traditabstract]{aa}
\usepackage{txfonts}
\usepackage{graphicx}
\usepackage{natbib}

\newcommand{\odp}{\mbox{\object{DP\,Leo}}}
\newcommand{\dpleo}{DP\,Leo}

\newcommand{\dpleoab}{DP\,Leo\,ab}
\newcommand{\dpleoabc}{DP\,Leo\,(ab)c}

\newcommand{\ten}[2]{#1\times 10^{#2}}
\newcommand{\msun}{M$_\odot$}
\newcommand{\mjup}{M$_\mathrm{Jup}$}

\begin{document}

\title{The giant planet orbiting the cataclysmic binary
  DP Leonis}

\author{
K. Beuermann\inst{1}
\and J. Buhlmann\inst{2}
\and J. Diese\inst{2}
\and S. Dreizler\inst{1}
\and F.~V. Hessman\inst{1}
\and T.-O. Husser\inst{1}
\and G.~F. Miller\inst{4}
\and N. Nickol\inst{2}
\and R. Pons\inst{2}
\and D. Ruhr\inst{2}
\and H. Schm\"ulling\inst{2}
\and A.~D. Schwope\inst{3}
\and T. Sorge\inst{2}
\and L. Ulrichs\inst{2}
\and D.~E. Winget\inst{4}
\and K.~I. Winget\inst{4}
}


\institute{
Institut f\"ur Astrophysik,
Georg-August-Universit\"at, Friedrich-Hund-Platz 1, 37077
G\"ottingen, Germany,
\and
Max-Planck-Gymnasium, Theaterplatz 10, 37073 G\"ottingen, Germany
\and Astrophysikalisches Institut Potsdam, An der Sternwarte 16, 14482
Potsdam, Germany
\and
Dept. of Astronomy, University of Texas at Austin,
RLM 16.236,  Austin, TX 78712, USA 
}

\date{Received 18 October 2010 / accepted 12 November 2010}

\authorrunning{K. Beuermann et al.} 
\titlerunning{The giant planet orbiting \dpleo}

\abstract 
{Planets orbiting post-common envelope binaries provide fundamental
  information on planet formation and evolution, especially for the
  yet nearly unexplored class of circumbinary planets. We searched for
  such planets in \odp, an eclipsing short-period binary, which
  shows long-term eclipse-time variations.  Using published,
  reanalysed, and new mid-eclipse times of the white dwarf in DP\,Leo,
  obtained between 1979 and 2010, we find agreement with the
  light-travel-time effect produced by a third body in an elliptical
  orbit. In particular, the measured binary period in 2009/2010 and
  the implied radial velocity coincide with the values predicted for
  the motion of the binary and the third body around the common center
  of mass. The orbital period, semi-major axis, and eccentricity of
  the third body are $P_\mathrm{c}\!=\!28.0\!\pm\!2.0$\,yrs,
  $a_\mathrm{c}\!=\!8.2\pm0.4$\,AU, and
  $e_\mathrm{c}\!=\!0.39\!\pm\!0.13$.  Its mass of
  sin\,$i_\mathrm{\,c}\,M_\mathrm{c}\!=\!6.1\!\pm\!0.5$\,\mjup\ qualifies
  it as a giant planet. It formed either as a first generation object
  in a protoplanetary disk around the original binary or as a second
  generation object in a disk formed in the common envelope shed by
  the progenitor of the white dwarf. Even a third generation origin in
  matter lost from the present accreting binary can not be entirely
  excluded. We searched for, but found no evidence for a fourth body.
}
{}{}{}{}


\maketitle

\section{Introduction}

Many eclipsing post-common envelope (CE) binaries, including the
cataclysmic variables, display long-term eclipse-time variations,
which represent either true or apparent changes of the orbital
period. True changes may result from the angular-momentum loss by
gravitational radiation or magnetic braking or from spin--orbit
exchange processes within the binary. Apparent changes may be effected
by apsidal motion or the presence of a third body. Eclipse-time
variations in cataclysmic variables (CVs) have often been attributed
to Applegate's (1992) mechanism of spin--orbit coupling, resulting
from changes in the internal constitution of the secondary, but this
process is generally too feeble to account for the observed amplitudes
\citep{brinkworthetal06,chen09,schwarzetal09}. Apsidal motion
\citep{todoran72} is unlikely to be present in CVs, because tidal
interaction is expected to circularize the orbits
effectively. Furthermore, apsidal motion can not account for the
observed non-sinusoidal shape of the eclipse-time variations
\citep[e.g.][]{beuermannetal10}. In recent years, interest has
therefore shifted back to the third-body hypothesis, which explains a
periodic variation of the eclipse times as the light-travel-time (LTT)
effect caused by the motion of the binary and an unseen third object
around the common center of gravity of the triple.

\odp\footnote{On recommendation by the Editor of A\&A, we refer to the
  system as DP\,Leo, to the binary explicitly as DP\,Leo\,ab and to
  the object orbiting the binary as \dpleoabc.} belongs to the still
small group of post-CE binaries known or suspected to possess planets,
among them HW\,Vir \citep{leeetal09}, NN\,Ser \citep{beuermannetal10},
HU\,Aqr \citep{schwarzetal09,nasirogluetal10}, and QS\,Vir
\citep{parsonsetal10}.  DP\,Leo\,ab is an 18.5 mag short-period
($P_\mathrm{orb}$\,=\,89.9\,min) polar, in which a synchronously
rotating magnetic white dwarf accretes matter from its Roche-lobe
filling companion \citep{schwopeetal02,pandeletal02}.  The observed
effective temperature of the white dwarf of 13\,500\,K
\citep{schwopeetal02} is likely due to accretional heating
\citep{townsleybildsten04}, suggesting an age exceeding the cooling
age of the white dwarf of 0.5\,Gyrs \citep{wood95}.

\citet{schwopeetal02} and \citet{pandeletal02} noted a decrease
of the binary period that could be described by a quadratic term in the
ephemeris. More recently, \citet{qianetal10} found that a reversal of
the long-term trend had taken place and suggested that the data
support a sinusoidal variation instead. They attributed this
modulation to a giant planet, which moves around the close binary in a
wide circular orbit with a period of 23.8 yrs.

The eclipsed light sources in \dpleo\ include the white dwarf, the
accretion spot on the white dwarf, and the magnetically controlled
accretion stream, which suffers only a grazing eclipse at the
inclination of $i\!=\!79.5^\circ$ \citep{schwopeetal02}. The part of
the stream closest to the white dwarf forms the accretion column,
which emits X-ray bremsstrahlung and optical cyclotron radiation,
heating the wider surrounding of the accretion spot by irradiation.
An accurate measurement of the period change in an accreting binary
requires that the contributions of these light sources to the observed
flux can be identified and the measured eclipse time can be reduced to
the mid-eclipse or superior conjunction of the white dwarf
\citep{schwopeetal02}. In early 2009, we started a long-term program
to measure accurate mid-eclipse times of the white dwarf in \dpleo.
This work is part of the ongoing effort of the University of
G\"ottingen to conduct research projects with high-school students.
 
\section{Observations and data analysis}

\begin{figure}[t]
\includegraphics[bb=169 46 547 700,height=88mm,angle=-90,clip]{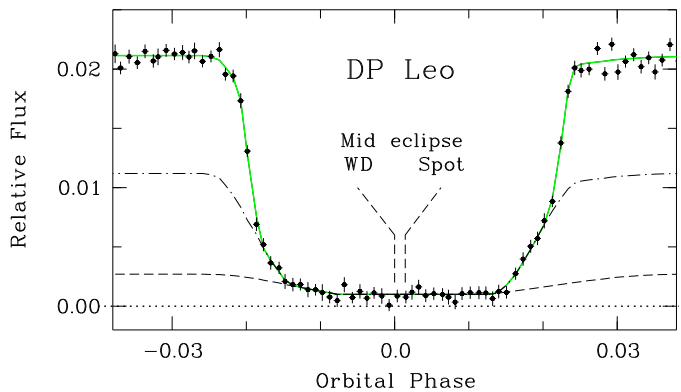}
\caption[chart]{Coadded light curve of 18 eclipses of \dpleo. The
  model curves represent the accretion stream (dashed), the sum of the
  stream component and the photospheric emission of the white dwarf
  (dot-dashed), and the sum of the latter and the spot component
  (solid).}
\label{fig:lc}
\end{figure}

Between 24 March 2009 and 19 February 2010, we measured 53 optical
eclipse light curves of \dpleo, using the remotely controlled
MONET/North 1.2-m telescope at McDonald Observatory via the MONET
internet remote-observing interface, either in white light or with a
Sloan r' filter. The exposure times were mostly 10\,s, separated by a
3\,s read-out interval.  Part of the observations were
taken by two groups of 10th and 11th grade high-school students at the
Max-Planck-Gymnasium G\"ottingen during normal school hours.
One additional eclipse was observed on 12 July 2010 with the McDonald
2.1-m telescope using a BG40 filter. We performed relative photometry
with respect to comparison star C1 (SDSS\,111722.63+175848.2), which
is located 31'' E and 65'' N of \dpleo\ and has Sloan r'$\,=\,$14.47
mag.

Throughout the observations, \dpleo\ showed the usual orbital
variation, reaching r'$\,\simeq\,$18 in the two cyclotron maxima and
dimming to r'$\,=\,$18.5--18.7 before eclipse and to r'$\,>\,$21 in
eclipse. In the first half of 2009, a noticeable flux from the
accretion stream was present, which almost vanished
later in the year and stayed low in 2010.
The white-dwarf eclipse in DP\,Leo is clearly discernible in light
curves measured with high sensitivity and time resolution. For the
MONET/North data, we obtained a sufficient S/N ratio by coadding 18
light curves taken in good atmospheric conditions in winter
2009/2010. Figure\,\ref{fig:lc} shows the resulting mean observed
light curve collected into phase bins of $\Delta \phi\!=\!0.001$. The
eclipse of the white dwarf and the accretion spot on the white dwarf
are clearly discernible at egress as the moderately and steeply rising
sections, respectively. The spot is seen near the trailing edge of the
white dwarf at egress and more centrally on the white dwarf at
ingress. The weak stream component contributes noticeably only around
$\phi\!\simeq\!-0.012$.  Employing information from our spring 2009
observations, when the stream was brighter, we modeled the stream
contribution in the individual eclipse light curves by a series of
concatenated straight lines. The average stream contribution is shown
by the dashed curve in Fig.\,\ref{fig:lc}. Our eclipse model
represents the white dwarf by a uniform disk and the excess emission
of the spot over the photospheric emission of the white dwarf by a
second smaller uniform disk. The latter is taken to include the
optical emission of the accretion column and of the heated polar cap
of the white dwarf. The assumed geometry of the spot is not relevant
for the present study. The combined fit of these contributions is
shown by the solid green curve in Fig.\,\ref{fig:lc}; the dot-dashed
black curve denotes the sum of the stream and the white dwarf. Before
least-squares fitting the model light curve, it was folded with the
10\,s exposure times. We find that the ingress and or egress of the
white dwarf takes $56.0\pm1.5$\,s, that of the spot last
$10.5\pm1.4$\,s. The eclipse of the accretion spot takes place $\Delta
t\!=\!7.6\,\pm\,1.5$\,s after superior conjunction of the white
dwarf. All quoted uncertainties refer to the unbiased 1-$\sigma$
errors obtained by stepping through the parameter in question with all
other parameters free.

Mid-eclipse times of the white dwarf were determined by least-squares
fitting the composite model to the individual light curves, using
$\Delta t\!=\!7.6$\,s and a relative flux of the white dwarf fixed at
the level shown in Fig.\,1. The fluxes of the spot at ingress and
egress and the linear functions that describe the time-dependent flux
of the stream were considered free parameters of the fit. We
determined the mid-eclipse time of the white dwarf and its error by
stepping the eclipse center in time and fitting a parabola to the
resulting $\,\chi^2$ variation. The derived white-dwarf mid-eclipse
times with their 1-$\sigma$ statistical errors are listed in
Table\,2. All times were shifted to the solar system barycenter,
corrected for leap seconds, and are quoted as Barycentric Julian Days
in the Terrestrial Time system. The 1.5-s uncertainty in $\Delta t$
represents an additional systematic error common to all eclipse times.

The 10th grade students used a simplified method of determining the
mid-eclipse times by visually cross-correlating the known eclipse
profile with the individual measured eclipse light curves. This
method, employed for didactic reasons, yielded internal errors for
individual eclipses of 3\,s, not much larger than those from the
formal fits used in this paper. 

\citet{schwopeetal02} measured two optical eclipses in white light
using the Optima high speed photometer on the Calar Alto 3.5-m
telescope in January 2002.  Of these, only the first was published.
We reanalysed the original data, which allowed us to discern the
eclipses of the white dwarf and the spot, employing the same model as
for the mean MONET light curve in Fig.\,1. The derived mid-eclipse
times of the white dwarf are given as the first two entries in
Table\,2. The 1-$\sigma$ formal errors of our fits are significantly
reduced over the conservative estimate of \citet{schwopeetal02} for
$E\!=\!56307$.

\citet{qianetal10} reported five eclipse times with a mean error of
6.9\,s. We did not include them in our analysis, because their
relation to the mid-eclipse of the white dwarf is uncertain and our
data cover the same time period with smaller errors.

\citet{schwopeetal02} summarized all eclipse-time measurements of
DP\,Leo available by 2002. Their list includes results from X-ray,
ultraviolet, and optical wavelengths, which they corrected to
represent the mid-eclipse time of the white dwarf, utilizing the known
slow secular drift of the accretion spot in azimuth. We adopt the 32
timings from \citet[][their Table\,2]{schwopeetal02} in addition to
the 56 timings from Table\,2. Our entire data set consists of 18
subsets loosely grouped in time, nine for the 1979-2002 data and nine
for our 2009/2010 measurements. Within each subset, statistics
dominate the scatter in the \mbox{$O\!-\!C$} eclipse-time variations.

\section{Results}

\begin{figure}[t]
\includegraphics[bb=170 39 547 699,height=88mm,angle=-90,clip]{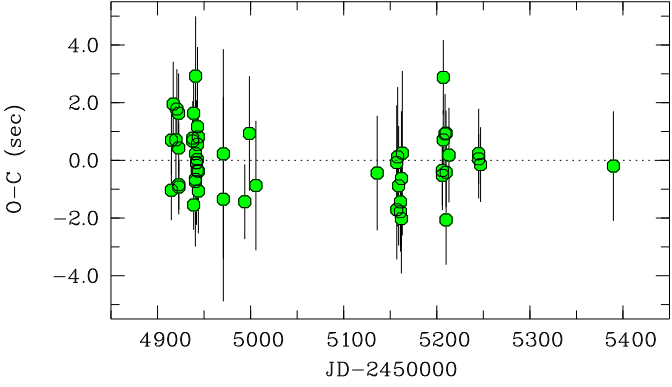}
\caption[chart]{Observed--Calculated time differences for the
  mid-eclipse of the white dwarf in DP Leo relative to the linear
  ephemeris of Eq.\,(1). }
\label{fig:linear}
\end{figure}

Our 54 mid-eclipse times of the white dwarf in \dpleo\ taken in
2009/2010 ($E\!=\!98482-106096$) define the linear ephemeris
\begin{equation}
\mathrm{BJD(TT)} = 2454914.8322920(20) + 0.06236285648(90)\,E,
\end{equation}
where the statistical errors are quoted in brackets and refer to the
last digits. The fit has a reduced $\,\chi^2_\nu\!=\!0.68$
($\,\chi^2\!=\!35.18$ for 52 d.o.f.). The residuals with respect to
the linear fit of Eq.\,(1) are given in Col.\,5 of Table\,2 and
displayed in Fig.\,\ref{fig:linear}.
The rms deviation from the fit is 1.1\,s. With 79\% of the values
deviating from the linear ephemeris by less than 1 $\sigma$, the
observed spread is consistent with a purely statistical origin. The
1.5-s error in $\Delta t$ represents an additional systematic error of
the first term of Eq.\,(1).

Including the pre-2002 white-dwarf mid-eclipse times of
\citet{schwopeetal02} yields an entirely different picture. While the
data available up to 2002 indicated a continuous period decrease
\citep{schwopeetal02,pandeletal02}, \citet{qianetal10} first noted
that a reversal had taken place some time after 2002 and suggested
that DP\,Leo exhibits a sinusoidal eclipse-time variation caused by a
third body in a circular orbit with a period of 23.8 yrs. Our larger
data set supports the third-body hypothesis by the detection of a
finite eccentricity of the orbit.

We fit the entire data set of 88 white-dwarf mid-eclipse times with the sum
of a linear ephemeris and the LTT effect produced by a third body
\dpleoabc, orbiting the binary \dpleoab,
\begin{equation}
T_\mathrm{ecl,WD}=T_0\,+\,P_\mathrm{bin}E + 
\frac{a_\mathrm{bin,c}\,\mathrm{sin}\,i_\mathrm{c}\,(1-e_\mathrm{c}^2)}{c\,(1+e_\mathrm{c}\,\mathrm{cos}\,\upsilon_\mathrm{bin,c})}\,\mathrm{sin}\,(\upsilon_\mathrm{bin,c}-\varpi_\mathrm{bin,c}),
\end{equation}
where $E$ is the cycle number. Note that we have not included the
quadratic term considered by \citet{schwopeetal02} and
\citet{pandeletal02}. The seven free parameters in the fit are the
epoch $T_0$, the binary period $P_\mathrm{bin}$, the amplitude of the
LTT effect
$K_\mathrm{bin,c}=\mathrm{sin}\,i_\mathrm{c}\,a_\mathrm{bin,c}/c$, the
orbital eccentricity $e_\mathrm{c}$, the longitude
$\varpi_\mathrm{bin,c}$ of the periastron from the ascending node in
the plane of the sky, the orbital period $P_\mathrm{c}$, and the
time $T_\mathrm{c}$ of periastron passage. The quantities
$i_\mathrm{c}$, $\upsilon_\mathrm{bin,c}$, $a_\mathrm{bin,c}$, and $c$
are the orbital inclination, the true anomaly, the semi-major axis of
the orbit of the center of mass of the binary around the common center
of mass of the triple, and the speed of light. The latter orbit is
point-symmetric with respect to that of body c and offset by $\pi$ in
longitude. Quantities referring to the binary orbit carry the index
'bin'. The motion of the center of gravity of the binary caused by the
third body is indicated by the index 'bin,c', and that of the third
body itself by 'c'. If the latter two quantities are identical we use
index 'c'.

\begin{figure}[t]
\includegraphics[bb=130 61 540 704,height=89mm,angle=-90,clip]{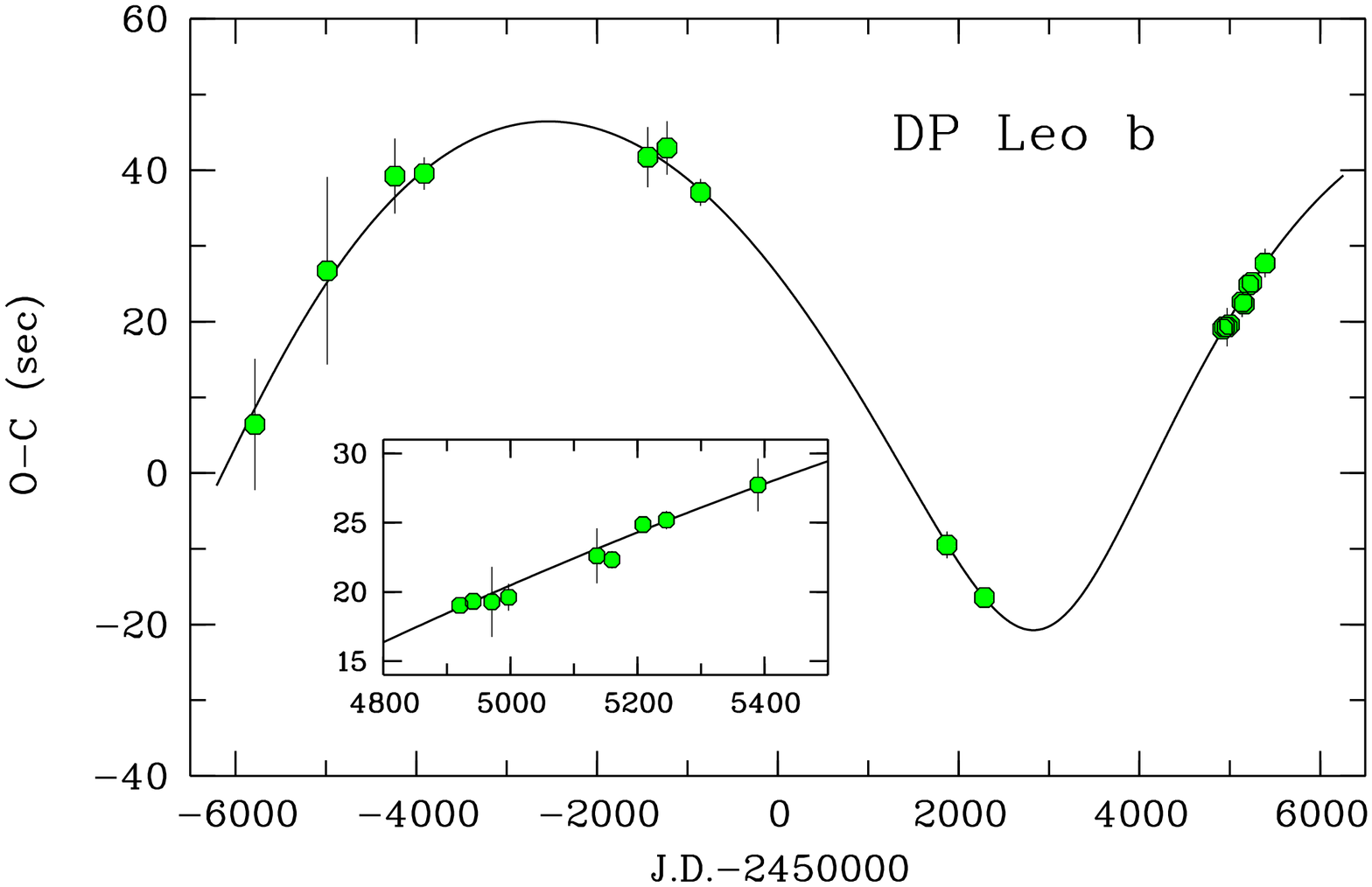}

\vspace{1mm}
\includegraphics[bb=327 61 538 704,height=89mm,angle=-90,clip]{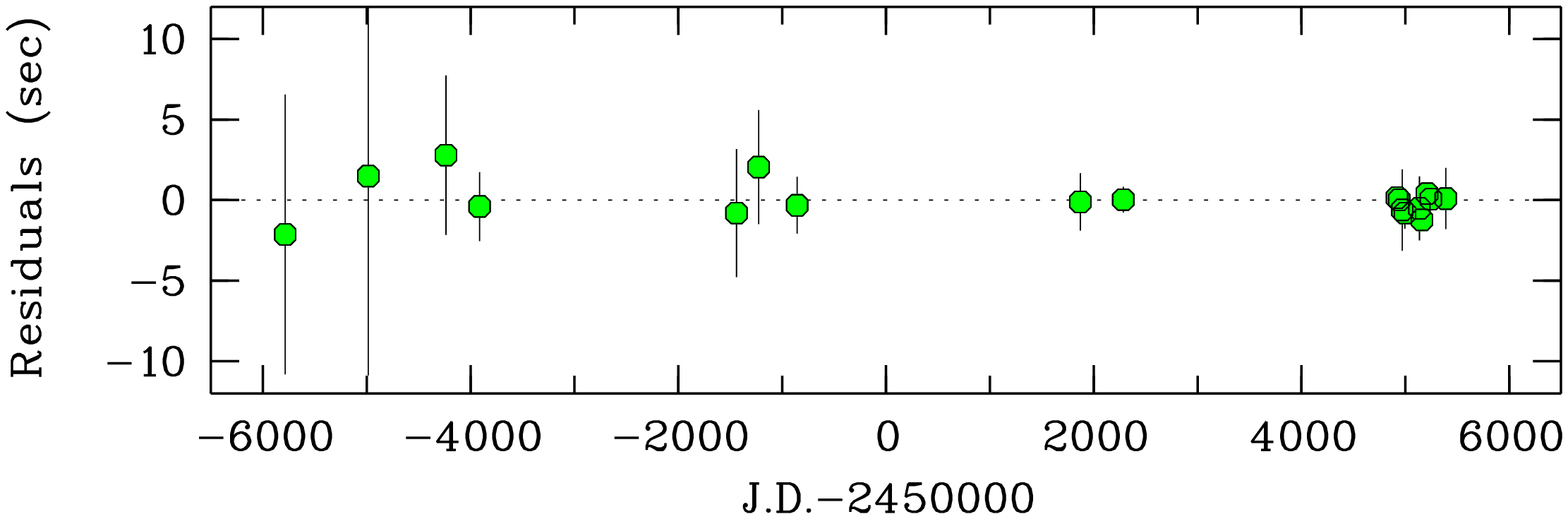}
\caption[chart]{Light-travel-time (LTT) effect produced by a third
  body in an elliptical orbit about \dpleo. \textit{Upper panel:
  }\mbox{$O\!-\!C$} time differences relative to a linear
  ephemeris. The curve shows the eccentric orbit fit, the data points
  represent the weighted means for 18 independent subsets of data
  taken between 1979 and 2010, as explained in the text. The inset is
  an enlargement of the section containing the new MONET/North
  data. \textit{Lower panel: } Residuals relative to the
  eccentric-orbit fit. }
\label{fig:1planet}
\end{figure}

The best fit to the 88 eclipse times yields an amplitude of the LTT
effect $K_\mathrm{bin,c}\!=\!33.7$\,s and a period
$P_\mathrm{c}\!=\!28.0$ yrs, with a reduced
$\,\chi^2_\mathrm{\nu}\!=\!0.77~(\,\chi^2\!=\!62.6$ for 81
d.o.f.). The orbit has an eccentricity
$e_\mathrm{c}\!=\!0.39\!\pm\!0.13$, with $e_\mathrm{c}\!=\!0$ excluded
at the 3.0\,$\sigma$ level. With
$\varpi_\mathrm{bin,c}\!=\!-78^\circ$, periastron occurs $12^\circ$
from the line of sight and was last passed near 2004.0.
Figure\,\ref{fig:1planet} shows the \mbox{$O\!-\!C$} eclipse-time
variations relative to the linear part of Eq.\,(2) in the upper panel
and the residuals from the fit including the third term of Eq.\,(2) in
the lower panel. While the curve represents the fit to all data
points, we avoid excessive clutter of the data points by displaying
only the weighted mean \mbox{$O\!-\!C$} values for each of the 18
subsets of the data, with the symbols placed at the weighted mean
eclipse times. The underlying time dependence of the \mbox{$O\!-\!C$}
variation becomes much clearer this way.
The parameters of \dpleoabc\ are summarized in Table\,1, which gives
also the unbiased 1-$\sigma$ errors determined by systematically
varying the parameter in question with all other parameters free. The
substantial correlated uncertainties in $\varpi_\mathrm{bin,c}$ and
$T_\mathrm{c}$ affect also the magnitude of the errors in the other
parameters and obliterate the small effect caused by the uncertainty
of the offset $\Delta t$ between the eclipses of white dwarf and
spot. For the first two entries in Table\,1, the errors are quoted in
brackets, referring to the last digit.

Assuming a central binary that consists of a white dwarf of
0.6\,\msun\ \citep{schwopeetal02} and a secondary of 0.1\,\msun, the
semi-major axis of the orbit of the third body becomes
$a_\mathrm{c}\!=\!8.2$\,AU. The amplitude $K_\mathrm{bin,c}$ of the
LTT effect implies an inclination-dependent mass
sin\,$i_\mathrm{c}\,M_\mathrm{c}\!=\!6.1$\,\mjup, qualifying the
object as a giant planet for any inclination
$i_\mathrm{c}\!>\!28^\circ$. The inclination $i_\mathrm{c}$ may be
close to the binary inclination of $79.5^\circ$, but can not be
determined by present means. The amplitude and the period of the
eclipse-time variations quoted by \citet{qianetal10} are roughly
consistent with the results presented here, while the sinusoidal form
of the eclipse-time variations assumed by them is not confirmed.

The $\,\chi^2$-minimization ensures that the eccentric-orbit fit meets
the centroid of our 2009/2010 data points. That the fit also
reproduces the observed derivative of the \mbox{$O\!-\!C$} curve
to a high degree of accuracy is not self-evident and provides
additional support for the third-body hypothesis. The eccentric-orbit
fit to all data points yields $P\!=\!0.0623628561$\,days for the
JD=\,2454919-2455389 time interval, in agreement with the observed period of
Eq.\,(1), $P\!=\!0.0623628565(9)$\,days. Both numbers refer to the
mean period in 2009/2010. Their agreement is illustrated in
the inset to Fig.\,3 and documented by the nearly identical residuals
for the linear  and eccentric-orbit fits in Table\,2. The
difference $\Delta P\!=\!P(t)-P_\mathrm{bin}$ between the 
period $P(t)$ at time $t$ and the intrinsic binary period
$P_\mathrm{bin}$ (Table\,1) varies between $-1.29$\,ms and $+1.52$\,ms
over the 28-year period. This range of $\Delta P$ corresponds to
radial velocities of the center of gravity of the binary on its path
around the center of gravity of the triple,
$v_\mathrm{rad}\!=\!\mathrm{c}\,\Delta P/P_\mathrm{bin}$, between
$-72$ and $+85$\,m\,s$^{-1}$. The observed period difference in
2009/2010, $\Delta P\!=\!1.20\!\pm\!0.08$\,ms, corresponds to
$v_\mathrm{rad}\!=\!66.8\!\pm\!4.5\,\mathrm{m\,s}^{-1}$ in agreement
with the velocity of $65$\,m\,s$^{-1}$ expected from the fit.

The present data provide a perfect fit without the quadratic term
$\frac{1}{2}\,P_\mathrm{bin}\dot{P}_\mathrm{bin}\,E^2$ that
\citet{schwopeetal02} and \citet{pandeletal02} included in the
ephemeris. This term measures the secular variation
$\dot{P}_\mathrm{bin}$ of the binary period. Its inclusion distorts
the run of the $O-C$ values displayed in Fig.\,3 by adding a parabolic
variation, which can be compensated for to some extent by adjusting
the fit parameters, notably $P_\mathrm{c}$ and $K_\mathrm{bin,c}$. The
fit deteriorates with increasing positive and negative values of
$\dot{P}_\mathrm{bin}$ and at the 1-$\sigma$ level, we find
$\dot{P}_\mathrm{bin}\!=\!\ten{(-0.3^{+1.0}_{-2.2})}{-14}$\,s\,s$^{-1}$,
about an order of magnitude lower than the negative value suggested by
\citet{schwopeetal02} and \citet{pandeletal02} and consistent with
zero. Obviously, a much longer time basis is needed to discern a
secular change of the binary period against the LTT effect produced by
the third body. The interpretation of the $O-C$ variation as the LTT
effect caused by a third body turns out to be robust.  However, a
finite $\dot{P}_\mathrm{bin}$, if it exists, would change some of the
fit parameters. For a negative $\dot{P}_\mathrm{bin}$ at the
1-$\sigma$ level in $\,\chi^2$, the planetary period, mass, and
eccentricity rise to $P_\mathrm{c}\!\simeq\!40$\,yrs,
sin\,$i_\mathrm{c}\,M_\mathrm{c}\!\simeq\!10$\,M$_\mathrm{Jup}$, and
$e_\mathrm{c}\!=\!0.45$.

\begin{table}[t]
\begin{flushleft}
\caption{Parameters derived from the eccentric orbit fit.
}
\begin{tabular}{l@{\hspace{7mm}}l}
\hline \\[-1ex]
Parameter & Value $\pm$ Error\\[1.0ex]
\hline\\[-1ex]
\multicolumn{2}{l}{\textit{(a) Binary DP\,Leo\,ab:}}\\[0.5ex]
Epoch $T_0$, BJD(TT) & 2448773.21461(9)   \\
Binary period $P_\mathrm{bin}$ (days) & 0.0623628426(6) \\[1.0ex]
\multicolumn{2}{l}{\textit{(b) Giant planet \dpleoabc:}}\\[0.5ex]
Orbital period $P_\mathrm{c}$ (years) & $28.0\pm2.0$\\
Semi-major axis $a_\mathrm{c}$ (AU) & $8.19\pm0.39$\\
Eccentricity $e_\mathrm{c}$ & $0.39\pm0.13$\\
Longitude of periastron $\varpi_\mathrm{bin,c}$ & $-78^\circ\pm20^\circ$ \\
Time of periastron passage $T_\mathrm{c}$, J.D. & $2453025\pm500$\\
Amplitude of the LTT effect $K_\mathrm{bin,c}$ (s) & $33.7\pm1.7$\\
Semi-major axis sin\,$i_\mathrm{c}\,a_\mathrm{bin,c}$ (cm) & $\ten{(1.01\pm0.05)}{12}$\\
Mass sin\,$i_\mathrm{c}\,M_\mathrm{c}$ (M$_\mathrm{Jup}$) & $6.05\pm0.47$   \\[1.5ex]
\hline\\[-3ex]
\end{tabular}
\label{tab:results}
\end{flushleft}
\end{table}

Detecting planets by the LTT effect is limited by the statistical and
systematic errors that affect the measurements of the mid-eclipse
times. Kepler's third law yields an LTT amplitude
$K_\mathrm{bin,c}\!=\!0.47\,M\,\mathrm{sin}\,i\,(P/M_\mathrm{bin})^{2/3}$\,s,
where $M_\mathrm{bin}$ is the binary mass in \msun, $M$ is the mass of
the planet in units of \mjup, $P$ its orbital period in years, and $i$
the inclination. The statistical and remaining systematic errors of
our 2009/2010 observations of DP\,Leo (Fig.\,2) may still hide a
modulation with an amplitude $\la\!0.5$\,s for a period of about
1\,year, allowing us to set an approximate upper limit to the mass of
an additional body with such a period of
$M\mathrm{sin}\,i\,\la\,0.8\,\mathrm{M}_\mathrm{Jup}$. Since the fits
yield $\,\chi^2_\mathrm{\nu}\!<\!1$, the present data provide no evidence
for a fourth body in DP\,Leo.

Systematic errors of the white-dwarf mid-eclipse times in polars arise
from the non-uniform brightness distribution of the white dwarf caused
by the accretion spot, its extended and variable structure, and the
formation of a heated pole cap by irradiation of the surrounding
photosphere. In DP\,Leo, the main accretion region forms a ribbon that
extends over about $30^\circ$ or $4\times10^8$\,cm on the surface of
the white dwarf \citep[][and references therein]{schwopeetal02},
accounting for the 10-s ingress and egress times of the spot
component. Weak emission has been detected from a second pole.  That
the mid-eclipse of the white dwarf in DP\,Leo can nevertheless be
measured to better than 1\,s is due to favorable circumstances: (i)
the accretion rate is sufficiently low to allow the identification of
the white-dwarf component in individual light curves; (ii) the
uneclipsed stream emission is faint, at least at times; and (iii) the
accretion geometry has remained unchanged since its discovery 1979. In
other polars, the accretion spot outshines the white dwarf, impeding
an easy reduction of the measured eclipse time to the mid-eclipse of
the white dwarf. The problem may be further aggravated if the white
dwarf is not synchronized, but rotates freely as in intermediate
polars. As a consequence, the 0.1\,s accuracy for the eclipse times of
the detached system NN\,Ser \citep{beuermannetal10} is probably not
attainable for such systems.

\section{Discussion}

The quality of our eccentric-orbit fit to the observed eclipse time
variations of DP\,Leo suggests that the detection of a third body is a
robust result. It can be tested by measuring the period (radial
velocity) evolution in the years to come, but the decisive next
periastron passage is expected to occur only in 2032.  Of the
alternative explanations, apsidal motion can easily produce the
observed amplitude for an eccentricity as small as
$e_\mathrm{bin}\!=\!0.01$ \citep{todoran72}, but predicts an
\mbox{$O\!-\!C$} variation that is sinusoidal to a high degree of
accuracy, and can therefore be excluded by the finite eccentricity
found by us. Applegate's (1992) mechanism, on the other hand, often
discussed in attempts to explain the observed eclipse-time variations
in CVs, can not produce the observed amplitude
\citep[e.g.][]{brinkworthetal06,chen09,schwarzetal09}. The detection
of planets in HW\,Vir \citep{leeetal09} and NN\,Ser
\citep{beuermannetal10}, as well as the likely detection in HU\,Aqr
\citep{schwarzetal09,nasirogluetal10} and QS\,Vir
\citep{parsonsetal10}, suggests that the occurrence of planets or
planetary systems in post-CE binaries may not be a rare incidence.

The progenitors of CVs are normal binaries with a primary of a couple
of solar masses and a low-mass secondary
\citep[e.g.][]{willemsetal05}. The mass of the primary is set by the
requirement that its core has reached the present white dwarf mass
when the star fills its Roche lobe, catastrophic mass transfer sets
in, and a CE is formed. Rapid spiral-in of the secondary leads to the
ejection of the envelope and the emergence of the newly born white
dwarf \citep[e.g.][]{sandquistetal98}. The system may become a CV if
angular momentum loss by gravitational radiation and magnetic braking
causes the secondary to reach its Roche lobe and mass transfer
resumes. In such systems, two principal paths of planet formation
exist: first generation planets that formed in a circumbinary
protoplanetary disk; and second generation planets that originated
from a disk formed in the ejected envelope \citep{perets10}. In the CE
phase, a pre-existing planet undergoes an outward motion caused by the
diminishing central mass and a less well understood inward motion
caused by the drag, which it experiences in the dense slowly expanding
envelope. This drag is usually ascribed to a supersonic Bondi-Hoyle
type momentum transfer \citep{alexanderetal76}. A preliminary study
suggests that the drag-induced inward drift may compensate for the
mass-related outward motion \citep[][and in
  preparation]{beuermannetal10}. This is particularly true if the
inclination of the planet is near that of the binary, ensuring that
the planet moves in the densest parts of the CE. The uncertainties
concerning the fate of a first generation planet are large, however,
and the origin of the planet in \dpleo\ must presently be considered
uncertain. Finally, there may be a third channel of planet formation
in matter lost from the secondary that was not accreted by the white
dwarf but left the binary and accumulated in a circumbinary
disk. Although such a disk may not contain enough mass for planet
formation \citep{taamspruit01}, the impact of a nova shell on the
stagnant matter could lead to the formation of planetesimals. Even
without such complications, planet formation around post-CE binaries
presents a variety of theoretical challenges.

\begin{acknowledgements}
This work is based on data obtained with the MONET telescopes funded
by the "Astronomie \& Internet" program of the Alfried Krupp von
Bohlen und Halbach-Foundation, Essen, and operated by the
Georg-August-Universit\"at G\"ottingen, the McDonald Observatory of
the University of Texas at Austin, and the South African Astronomical
Observatory, on data obtained with the 2.1-m telescope at McDonald
Observatory, and on data obtained with the 3.5-m telescope of the
German-Spanish Astronomical Centre, Calar Alto. We acknowledge helpful
comments by the anonymous referee. We also thank Andreas Seifahrt and
Ulf Seemann for taking part of the light curves with the MONET/North
telescope. We gratefully acknowledge the support from the
Robert-Bosch-Foundation by awarding the Robert-Bosch-Prize 2010
``Schule trifft Wissenschaft'' to our collaborative project.
\end{acknowledgements}

\begin{table}[thb]
\begin{flushleft}
\caption{Reanalysed and new white-dwarf mid-eclipse times of DP Leo.
}
\begin{tabular}{c@{\hspace{4mm}}c@{\hspace{4mm}}c@{\hspace{3mm}}c@{\hspace{5mm}}c@{\hspace{2mm}}c}
\hline \\[-1ex]
Cycle No. & BJD(TT) &  \multicolumn{2}{c}{Error} & \multicolumn{2}{c}{Residuals (s)}\\
$E$& 2400000+ & (days) & (s) & linear & eccentric\\[0.5ex]
\hline\\[-1ex]
\multicolumn{6}{l}{\textit{(a) Calar Alto 3.5-m, Optima white-light photometry, reanalysed}} \\[0.5ex]
  56307 & 52284.678997 & 0.000011 & 0.95 & -- & \hspace{-2mm}$-$0.05 \\
  56308 & 52284.741363 & 0.000017 & 1.44 & -- & 0.22 \\[0.5ex]
\multicolumn{6}{l}{\textit{(b) MONET/North 1.2-m, white-light photometry}} \\[0.5ex]
  98482 & 54914.832280 & 0.000012 & 1.04 & \hspace{-2mm}$-$1.04 & \hspace{-2mm}$-$1.01 \\
  98483 & 54914.894663 & 0.000014 & 1.21 &  0.70 &  0.73 \\
  98514 & 54916.827926 & 0.000017 & 1.47 &  1.95 &  1.97 \\
  98560 & 54919.696603 & 0.000017 & 1.47 &  0.71 &  0.73 \\
  98577 & 54920.756784 & 0.000016 & 1.38 &  1.78 &  1.80 \\
  98607 & 54922.627668 & 0.000016 & 1.38 &  1.63 &  1.65 \\
  98608 & 54922.690017 & 0.000010 & 0.86 &  0.43 &  0.46 \\
  98609 & 54922.752365 & 0.000010 & 0.86 & \hspace{-2mm}$-$0.85 & \hspace{-2mm}$-$0.83 \\
  98610 & 54922.814727 & 0.000011 & 0.95 & \hspace{-2mm}$-$0.92 & \hspace{-2mm}$-$0.91 \\
  98850 & 54937.781831 & 0.000016 & 1.38 &  0.67 &  0.67 \\
  98851 & 54937.844195 & 0.000012 & 1.04 &  0.77 &  0.77 \\
  98865 & 54938.717285 & 0.000015 & 1.30 &  1.63 &  1.63 \\
  98866 & 54938.779611 & 0.000010 & 0.86 & \hspace{-2mm}$-$1.55 & \hspace{-2mm}$-$1.55 \\
  98896 & 54940.650507 & 0.000027 & 2.33 & \hspace{-2mm}$-$0.66 & \hspace{-2mm}$-$0.66 \\
  98897 & 54940.712869 & 0.000011 & 0.95 & \hspace{-2mm}$-$0.73 & \hspace{-2mm}$-$0.73 \\
  98898 & 54940.775243 & 0.000028 & 2.42 &  0.23 &  0.22 \\
  98899 & 54940.837637 & 0.000024 & 2.07 &  2.92 &  2.92 \\
  98914 & 54941.773045 & 0.000014 & 1.21 & \hspace{-2mm}$-$0.09 & \hspace{-2mm}$-$0.10 \\
  98915 & 54941.835408 & 0.000025 & 2.16 & \hspace{-2mm}$-$0.08 & \hspace{-2mm}$-$0.08 \\
  98928 & 54942.646122 & 0.000011 & 0.95 & \hspace{-2mm}$-$0.35 & \hspace{-2mm}$-$0.35 \\
  98930 & 54942.770852 & 0.000012 & 1.04 &  0.03 &  0.02 \\
  98931 & 54942.833221 & 0.000012 & 1.04 &  0.55 &  0.55 \\
  98932 & 54942.895591 & 0.000032 & 2.76 &  1.17 &  1.17 \\
  98945 & 54943.706304 & 0.000010 & 0.86 &  0.81 &  0.81 \\
  98946 & 54943.768653 & 0.000012 & 1.04 & \hspace{-2mm}$-$0.38 & \hspace{-2mm}$-$0.39 \\
  98947 & 54943.831008 & 0.000017 & 1.47 & \hspace{-2mm}$-$1.06 & \hspace{-2mm}$-$1.07 \\
  99377 & 54970.647033 & 0.000041 & 3.54 & \hspace{-2mm}$-$1.35 & \hspace{-2mm}$-$1.37 \\
  99378 & 54970.709414 & 0.000042 & 3.63 &  0.22 &  0.19 \\
  99746 & 54993.658926 & 0.000015 & 1.30 & \hspace{-2mm}$-$1.43 & \hspace{-2mm}$-$1.49 \\
  99826 & 54998.647982 & 0.000023 & 1.99 &  0.93 &  0.89 \\
  99938 & 55005.632601 & 0.000026 & 2.25 & \hspace{-2mm}$-$0.87 & \hspace{-2mm}$-$0.92 \\
 102028 & 55135.970976 & 0.000023 & 1.99 & \hspace{-2mm}$-$0.44 & \hspace{-2mm}$-$0.51 \\
 102364 & 55156.924900 & 0.000023 & 1.99 & \hspace{-2mm}$-$0.08 & \hspace{-2mm}$-$0.13 \\
 102365 & 55156.987244 & 0.000020 & 1.73 & \hspace{-2mm}$-$1.71 & \hspace{-2mm}$-$1.76 \\
 102380 & 55157.922708 & 0.000028 & 2.42 &  0.12 &  0.07 \\
 102397 & 55158.982865 & 0.000024 & 2.07 & \hspace{-2mm}$-$0.88 & \hspace{-2mm}$-$0.93 \\
 102428 & 55160.916107 & 0.000018 & 1.56 & \hspace{-2mm}$-$1.44 & \hspace{-2mm}$-$1.49 \\
 102429 & 55160.978466 & 0.000016 & 1.38 & \hspace{-2mm}$-$1.78 & \hspace{-2mm}$-$1.83 \\
 102444 & 55161.913906 & 0.000022 & 1.90 & \hspace{-2mm}$-$2.02 & \hspace{-2mm}$-$2.07 \\
 102445 & 55161.976285 & 0.000027 & 2.33 & \hspace{-2mm}$-$0.63 & \hspace{-2mm}$-$0.68 \\
 102460 & 55162.911738 & 0.000033 & 2.85 &  0.25 &  0.20 \\
 103150 & 55205.942102 & 0.000014 & 1.21 & \hspace{-2mm}$-$0.35 & \hspace{-2mm}$-$0.37 \\
 103151 & 55206.004463 & 0.000014 & 1.21 & \hspace{-2mm}$-$0.52 & \hspace{-2mm}$-$0.53 \\
 103166 & 55206.939920 & 0.000012 & 1.04 &  0.71 &  0.69 \\
 103167 & 55207.002308 & 0.000015 & 1.30 &  2.88 &  2.87 \\
 103197 & 55208.873171 & 0.000016 & 1.38 &  0.92 &  0.91 \\
 103213 & 55209.870977 & 0.000009 & 0.78 &  0.94 &  0.93 \\
 103214 & 55209.933305 & 0.000018 & 1.56 & \hspace{-2mm}$-$2.06 & \hspace{-2mm}$-$2.08  \\
 103215 & 55209.995687 & 0.000014 & 1.21 & \hspace{-2mm}$-$0.41 & \hspace{-2mm}$-$0.42  \\
 103263 & 55212.989111 & 0.000019 & 1.64 &  0.18 &  0.17 \\
 103775 & 55244.918892 & 0.000010 & 0.86 &  0.05 &  0.08 \\
 103776 & 55244.981257 & 0.000018 & 1.56 &  0.23 &  0.27 \\
 103808 & 55246.976864 & 0.000015 & 1.30 & \hspace{-2mm}$-$0.15 & \hspace{-2mm}$-$0.11 \\[0.5ex]
\multicolumn{6}{l}{\textit{(c) McDonald 2.1-m, photometry with BG\,40 filter}} \\[0.5ex]
 106096 & 55389.664062 & 0.000022 & 1.90 & \hspace{-2mm}$-$0.20 & 0.10 \\[0.5ex]
\hline\\[-7ex]
\end{tabular}
\label{tab:data}
\end{flushleft}
\end{table}

\bibliographystyle{aa}

\end{document}